\documentclass[twocolumn,pra,superscriptaddress,notitlepage]{revtex4-1}

\usepackage{amsmath}
\usepackage{graphicx} 
\usepackage{here}

\usepackage{amsmath,amssymb,amsthm,mathrsfs,amsfonts,dsfont}
\usepackage{braket}
\usepackage{bm}
\usepackage{enumerate}
\usepackage{color}
\usepackage{algorithm}
\usepackage{comment}%

\usepackage{amsmath,amssymb,amsthm,mathrsfs,amsfonts,dsfont}
\usepackage{braket}
\usepackage{bm}
\usepackage{enumerate}
\usepackage{color}
\usepackage{qcircuit}
\usepackage{amsmath,bm}
\usepackage{braket}
\usepackage{bm}
\usepackage{enumerate}
\usepackage{color}
\usepackage{graphicx}
\usepackage{algpseudocode}
\usepackage{comment}%
\newcommand{\nn}{\notag \\}

\newcommand{\tr}{\mathrm{Tr}}%
\newcommand{\hH}{\hat{H}}%
\newcommand{\hU}{\hat{U}}

\newcommand{\hsig}{\hat{\sigma}}
\newcommand{\hrho}{\hat{\rho}}

\begin{document}


\title{Spectroscopic estimation of the photon number for  superconducting \newline
Kerr parametric oscillators}


\author{Keisuke Matsumoto  }
\email{1221544@ed.tus.ac.jp, matsumoto-kei@aist.go.jp}
\affiliation{Department of Physics, Tokyo University of Science,
1-3, Kagurazaka, Shinjuku-ku, Tokyo, 162-8601, Japan.}
\affiliation{Research Center for Emerging Computing Technologies, National Institute of Advanced Industrial Science and Technology (AIST), Umezono1-1-1, Tsukuba, Ibaraki 305-8568, Japan.}

 \author{Aiko Yamaguchi}
 \affiliation{Secure System Platform Research Laboratories, NEC Corporation, 
1753, Shimonumabe, Kawasaki, Kanagawa 211-0011, Japan}
 \affiliation{NEC-AIST Quantum Technology Cooperative Research Laboratory, National Institute of Advanced Industrial Science and Technology (AIST), Tsukuba, Ibaraki 305-8568, Japan}

  \author{\ \ \ \ \ \ \ \ \ \\ \ \  \ \ \ Tsuyoshi Yamamoto}
 \affiliation{Secure System Platform Research Laboratories, NEC Corporation, 
1753, Shimonumabe, Kawasaki, Kanagawa 211-0011, Japan}
 \affiliation{NEC-AIST Quantum Technology Cooperative Research Laboratory, National Institute of Advanced Industrial Science and Technology (AIST), Tsukuba, Ibaraki 305-8568, Japan}

 \author{Shiro Kawabata}
 \affiliation{Research Center for Emerging Computing Technologies, National Institute of Advanced Industrial Science and Technology (AIST), Umezono1-1-1, Tsukuba, Ibaraki 305-8568, Japan.}
 \affiliation{NEC-AIST Quantum Technology Cooperative Research Laboratory, National Institute of Advanced Industrial Science and Technology (AIST), Tsukuba, Ibaraki 305-8568, Japan}

\author{Yuichiro Matsuzaki 
}
\email{matsuzaki.yuichiro@aist.go.jp}  
\affiliation{Research Center for Emerging Computing Technologies, National Institute of Advanced Industrial Science and Technology (AIST), Umezono1-1-1, Tsukuba, Ibaraki 305-8568, Japan.}
\affiliation{NEC-AIST Quantum Technology Cooperative Research Laboratory, National Institute of Advanced Industrial Science and Technology (AIST), Tsukuba, Ibaraki 305-8568, Japan}


\begin{abstract}
Quantum annealing (QA) is a way to solve combinational optimization problems. Kerr nonlinear parametric oscillators (KPOs) are promising devices for implementing QA. 
When we solve the combinational optimization problems using KPOs, it is necessary to precisely control the photon number of the KPOs.
Here, we propose a feasible method to estimate the photon number of the KPO. We consider coupling an ancillary qubit to the KPO and show that spectroscopic measurements on the ancillary qubit provide information on the photon number of the KPO.
\end{abstract}

\maketitle

\section{Introduction}
Recently, 
much attention has been paid to Kerr nonlinear parametric oscillators (KPOs)~\cite{goto2016bifurcation,goto2019quantum,goto2019demand,wang2019quantum,grimm2020stabilization,yamaji2022spectroscopic}.
KPO is based on Kerr-nonlinear resonators
driven by two-photon excitation
~\cite{PhysRevA.44.4704,PhysRevA.48.2494},
which can be realized by using
superconducting resonators~\cite{bourassa2012josephson,meaney2014quantum,leghtas2015confining}.
It is known that the KPO can be used as a qubit for a gate-type quantum computer
~\cite{cochrane1999macroscopically,goto2016universal,puri2017engineering,puri2020bias}.
Recently, it has been shown that we can observe quantum phase transitions of
the KPOs, which is useful for quantum metrology~\cite{bartolo2016exact,minganti2018spectral,dykman2018interaction,rota2019quantum}.

Also, KPO is a promising
candidate for realizing quantum annealing (QA)~\cite{goto2016bifurcation,puri2017quantum}.
Quantum annealing (QA) is one of the techniques to solve combinational optimization problems~\cite{kadowaki1998quantum,farhi2001quantum}. The solution to the problems can be embedded into a ground state of the Ising Hamiltonian, and we can
obtain the ground state of the Ising Hamiltonian
after performing QA as long as the adiabatic condition is satisfied \cite{morita2008mathematical}. 

Importantly, the Hamiltonian of the KPOs can be mapped into an Ising Hamiltonian ~\cite{goto2016bifurcation,puri2017quantum}.
To implement QA with KPOs, we start from vacuum states, and we gradually increase parametric driving terms in an adiabatic way. Then,
the network of the KPOs finds a ground state of the Hamiltonian via a bifurcation process.
A feasible architecture for QA with KPOs 
using nearest neighbor interactions has also been proposed~\cite{lechner2015quantum,puri2017quantum}.

However, in order to accurately map the Ising Hamiltonian to the KPO Hamiltonian, we need to precisely control the average number of photons of each KPO. Although there is a formula to calculate the number of photons of the KPO under a semi-classical approximation, the calculated value can be different from the actual value~\cite{kanao2021high}. So a reliable way to estimate the number of the photons of the KPO is required to solve practical combinational optimization problems.

\begin{figure}[h]
\centering
		\includegraphics[width=9.1cm]{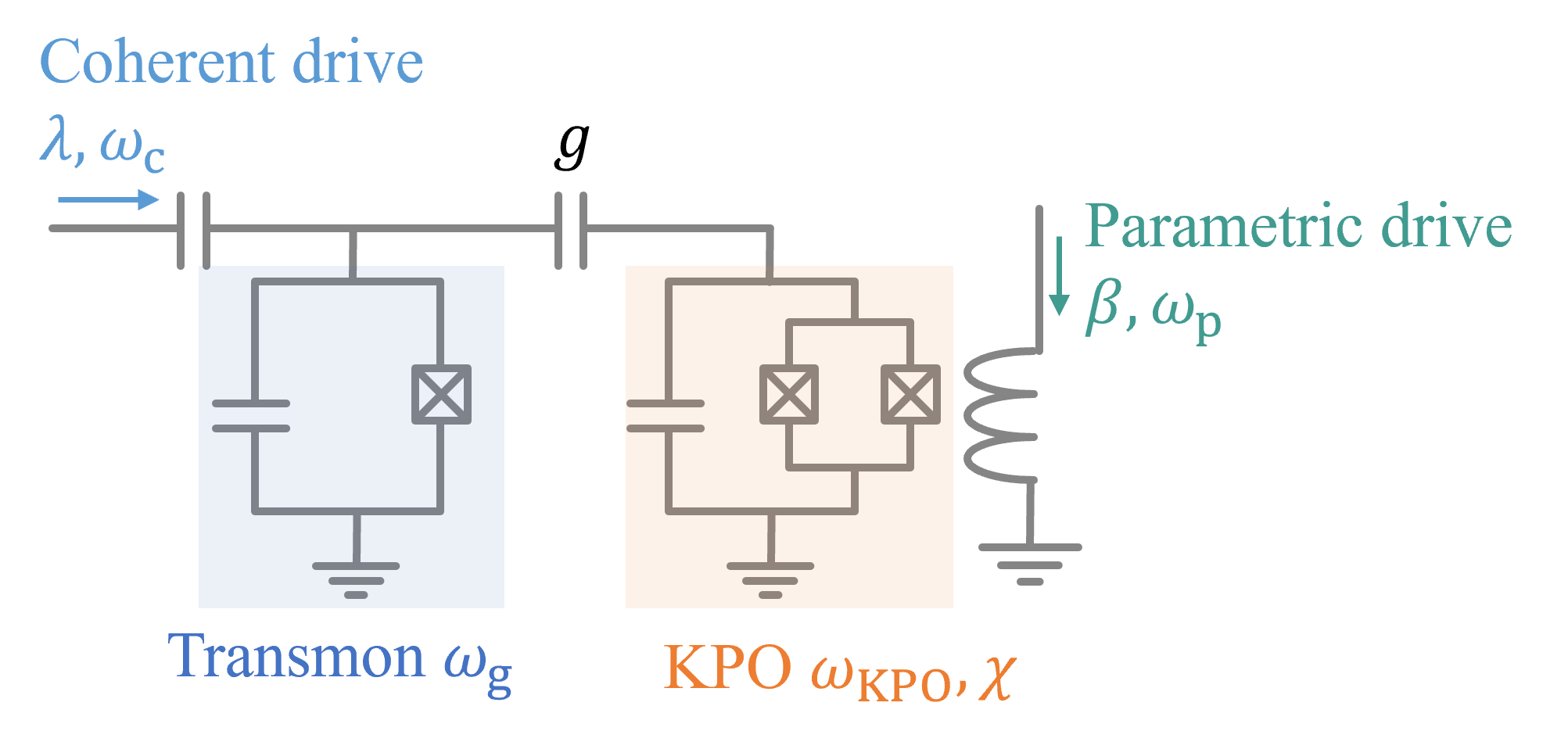} \\
\caption{Schematic of a KPO coupled with a transmon qubit.}
\label{fig:schematic1}
\end{figure}

In this paper, we propose a method to estimate the number of photons of the KPO from spectroscopic measurement. We consider a system, where the KPO is coupled with an ancillary qubit such as a superconducting transmon qubit or another KPO (without parametric drive), as shown in Fig.~\ref{fig:schematic1}. We show that  spectroscopic measurements on the ancillary qubit provide an estimate of the number of photons of the KPO. We evaluate the performance of our method with numerical simulations by solving a master equation and show that the proposed method is more accurate than the conventional method.

The paper is organized as follows. In Sec. II, we introduce a model of a KPO coupled with an ancillary qubit. In Sec. III, we describe our method to estimate the number of the photons of the KPO by spectroscopic measurements.
In Sec. IV, we evaluate the performance of our method by using numerical simulations.
In Sec. V, we conclude our discussion.
Throughout this paper, we set $\hbar=1$.

\section{
MODEL HAMILTONIAN
}\label{Model}
In this section, we introduce a model of a KPO coupled 
with an ancillary qubit.
The Hamiltonian is given by
\begin{align}
    \hH&=\omega_{\rm{KPO}}\ \hat{a}^\dagger\hat{a} - \frac{\chi}{12} (\hat{a}+\hat{a}^\dagger)^4 + 2\beta (\hat{a} + \hat{a}^{\dagger})^2\cos{\omega_p t}\nn[10pt]
    &+\frac{\omega_{\rm{g}}}{2}\hsig_z
    +g(\hat{a}+\hat{a}^\dagger)\hsig_x
    +\lambda \hsig_x \cos{\omega_c t},
\end{align}
where $\hat{a}^\dagger$ ($\hat{a}$) is a creation
(annihilation) operator of the KPO, $\omega_{\rm{KPO}}$ is the frequency of the KPO, $\chi$ is the Kerr coefficient, $\beta$ is the amplitude of a parametric drive, $\omega_p$ is the frequency of the
parametric drive, $\omega_{\rm{g}}$ is the frequency of the ancillary qubit, $g$ is the coupling 
strength between the
KPO and the ancillary qubit, 
and $\lambda$ $(\omega_c)$ is the amplitude (frequency) of the driving field for the qubit, respectively. Here, $\hsig_x$ and $\hsig_z$ denote the Pauli operators.
Moving into a rotating frame at the frequency of $\omega_p/2$ and adapting the rotating wave approximation, the Hamiltonian is written as
\begin{align}\label{total hamiltonian}
  \hH&=\hH_{\rm{KPO}}+\hH_{\rm{G}}+\hH_{\rm{I}}+\hH_{\rm{D}},\\[10pt]
    \hH_{\rm{KPO}}&=\Delta \hat{a}^\dagger\hat{a} - \frac{\chi}{2} \hat{a}^\dagger\hat{a}^\dagger\hat{a}\hat{a} + \beta (\hat{a}^2 + \hat{a}^{\dagger 2})\\[10pt]
    \hH_{\rm{G}}&=\frac{\omega_{\rm{g}}-\omega_{\rm{p}}/2}{2}\hsig_z\\[10pt]
    \hH_{\rm{I}}&=g(\hat{a}\hsig_{+}+\hat{a}^\dagger\hsig_{-})\\[10pt]
    \hH_{\rm{D}}&=\lambda_{\rm{p}}\left(\hsig_+ e^{-i(\omega_c-\omega_p/2)t}+\hsig_- e^{i(\omega_c-\omega_p/2)t}\right),
\end{align}
where $\Delta = \omega_{\rm{KPO}} -\chi - \omega_p/2$ denotes the detuning of the KPO,  $\lambda_{\rm{p}}={\lambda}/{2}$ denotes the Rabi frequency of the ancillary qubit, and $\hat{\sigma}_{\pm}$ denotes the ladder operator.
Throughout our paper, we set $\Delta <0$.
The ground and the first excited states of $\hH_{\rm{G}}$ are $|{\rm{g}}\rangle $ and $|{\rm{e}}\rangle $, respectively. With $\beta =0$, 
the Fock states
$\ket{n}$ (for $n=0,1,2, 3$) become eigenstates of the
$H_{\rm{KPO}}$. For $\beta \gg |\chi|$, on the other hand, the corresponding eigenstates are approximately given by $(|\alpha \rangle+ |-\alpha \rangle )/\sqrt{2}$, $(|\alpha \rangle- |-\alpha \rangle )/\sqrt{2}$, $({\it{D}}_{\alpha}+{\it{D}}_{-\alpha})|1\rangle /\sqrt{2}$, and 
$({\it{D}}_{\alpha}-{\it{D}}_{-\alpha})|1\rangle /\sqrt{2}$, where ${\it{D}}_{ \alpha}=\exp{(\alpha\hat{a}-\alpha^{\ast}\hat{a}^\dagger)}$
denotes a displacement operator~\cite{goto2016bifurcation}.

\section{Methods}
In this section, we propose a method to estimate the number of photons of the KPO from a spectroscopic measurement of an ancillary qubit coupled with the KPO. As we explained,
for sufficiently large $\beta$, the ground state of the KPO is approximately described by a superposition of two coherent states, namely $|\alpha\rangle$ and $|-\alpha\rangle$ where $\pm \alpha$ is the amplitude of the coherent state. Without loss of generality, we can assume that $\alpha$ is a real number. Then, the Hamiltonian of the ancillary qubit is approximately written as

\begin{align}\label{qubit_Hamiltonian}
    \hH_{\rm{qubit}}^{\rm{eff}}
    &=\frac{\omega_{\rm{g}}-\omega_{\rm{p}}/2}{2}\hsig_z
    \pm 
    g\alpha\ \hsig_x\nn[10pt]
    &+\lambda_{\rm{p}}\left(\hsig_+ e^{-i\Delta _{\rm{q}}t}+\hsig_- e^{i\Delta _{\rm{q}}t}\right),
\end{align}
where $\Delta _{\rm{q}}=\omega_c-\omega_p/2$ denotes a detuning of the ancillary qubit \cite{goto2016bifurcation}.

It is known that we can observe a Mollow triplet via a spectroscopic measurement with this Hamiltonian, where resonant transition frequencies are 
$\Delta _{\rm{q}}=0$, $\pm g\alpha$~\cite{Mollow:1969zz,wu1994phase,wrigge2008efficient,ulhaq2012cascaded,xu2007coherent,laucht2017dressed}.
Since we can estimate the value of $g$ from a separate spectroscopic measurement by observing a vacuum Rabi splitting \cite{wallraff2004strong,chiorescu2004coherent}, 
we can obtain the value of $\alpha$ from the peak (dip) positions observed in the Mollow triplet.

With a conventional method \cite{puri2017engineering}, an analytical formula under semi-classical approximations such as $\alpha_{\rm{ana}}=\sqrt{(2\beta+\Delta)/\chi}$ is used to estimate the number of photons of the KPO \footnote{Strictly speaking, a decay rate $\gamma_1$ also affects the number of photons. However, since we use parameters $\beta$ which is much larger than the decay rate, we can ignore the effect of $\gamma_1$ for the estimation of the number of photons \cite{puri2017engineering}}. However,
previous research shows that this value can provide a wrong estimate~\cite{kanao2021high} due to the violation of the approximation.

\section{Numerical simulations}
In this section, we evaluate the performance of our method by comparing with the conventional method using numerical simulations of the GKSL (Gorini-Kossakowski-Sudarshan-Lindblad) master equation.
Here, we adopt the Hamiltonian in Eq.~\eqref{total hamiltonian}.
To take the effect of decoherence into account, we use the following GKSL master equation

\begin{align}\label{Eq.GKSL}
\frac{\partial\rho}{\partial t} &= -i\left[\hH_{\rm{KPO}}+\hH_{\rm{G}}+\hH_{\rm{I}}+\hH_{\rm{D}} 
,\ \rho\right] \nn[10pt]
&+\frac{\gamma_1}{2} \left (2\hat a \rho \hat a^\dagger - \left\{ \hat a^\dagger \hat a, \rho \right\}\right)
+ \frac{\gamma_2}{2} \left (2\hsig_{-} \rho \hsig_{+} - \left\{ \hsig_{+} \hsig_{-}, \rho \right\}\right),
\end{align}
where $\gamma_1$ denotes the one photon dissipation rate of the KPO, $\gamma_2$ denotes the spontaneous emission rate of the ancillary qubit, and
$\hrho$ denotes the density matrix describing the quantum state of the total system.
We solve the GKSL master equation Eq.~\eqref{Eq.GKSL} using QuTiP~\cite{johansson2012qutip}.
We choose the initial state as a
steady state of Eq.~\eqref{Eq.GKSL} with $\lambda _{\rm{p}}=0$.

\begin{figure}[h]
\centering
		\includegraphics[width=9.1cm]{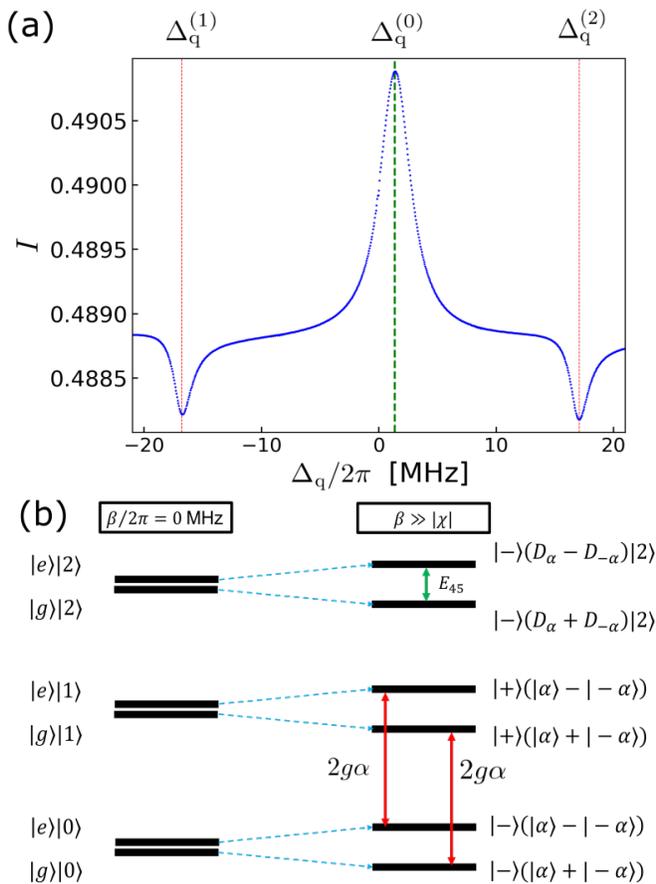} \\
\caption{
(a) The time-integrated spectra $I$ against $\Delta_{\rm{q}}/2\pi$ with $\lambda_{\rm{p}}/2\pi=0.5$  MHz. We set the parameters as $\Delta/2\pi=-30.0$  MHz, $\chi/2\pi=18.0$ MHz, $\beta/2\pi=42.0$ MHz, $g/2\pi=5.0$ MHz, $\gamma_1/2\pi=\gamma_2/2\pi=0.8$ MHz, and $\omega_g=\omega_p/2$.
(b)The energy diagram of the states of a KPO coupled with a qubit.
In the left (right) side, we show the energy diagram with $\beta/2\pi=0$ MHz ($\beta\gg |\chi|$). Here,
$|\alpha \rangle $ and $\ket{n}$ (for $n=0,1,2$) denote a coherent state
 and Fock states, respectively, while
${\it{D}}_{ \alpha}=\exp{(\alpha\hat{a}-\alpha^{\ast}\hat{a}^\dagger)}$
denotes the discplacement operator. 
Also, $|g\rangle $ ($|e\rangle $) and $|\pm \rangle =\frac{1}{\sqrt{2}}(|g\rangle \pm |e\rangle )$
denotes the ground (excited) state and the superposition states of the qubit.
}
\label{fig:spect1}
\end{figure}

In Fig.~\ref{fig:spect1} (a), we plot a time-integrated spectra $I=(1/t)\int_{0}^{t}\ d\tau\ (\braket{\hsig_z}+1)/2$ as a function of $\Delta_{\rm{q}}$ with a step of $0.05$ MHz,
where $\braket{\hsig_z}={\rm{Tr}}[\hsig_z \rho ]$,
which is effectively the same as a spectroscopy to detect the change in the pupulation of the qubit.

This spectra is upper (lower) bounded by $1$ ($0$).
The value of this spectra depends on the Rabi frequency and decay rate of the qubit.
The observed peak and dips are at $\Delta_{\rm{q}}^{(0)}/2\pi=1.35$ MHz, $\Delta_{\rm{q}}^{(1)}/2\pi= -16.80$ MHz and $\Delta_{\rm{q}}^{(2)}/2\pi = 17.10$ MHz, respectively.

Fig.~\ref{fig:spect1} (b) shows the energy diagram composed of the system of the KPO coupled with an ancillary qubit.
We calculate the energy eigenvalues of the Hamiltonian, and we confirm that the energy difference between the eigenvalues is almost the same as the peak frequency observed in our numerical simulation.
The dip at $\Delta_{\rm{q}}^{(1)}/2\pi= -16.80$ MHz ($\Delta_{\rm{q}}^{(2)}/2\pi = 17.10$ MHz) corresponds to the transition between the ground (first excited) state and the second (third) excited state, which we describe by a red vertical arrow in Fig.~\ref{fig:spect1} (b).
Here, with our parameters, the ground state, first excited state, second excited state, and third excited state 
 are approximately described as
$\ket{-}(\ket{\alpha}+\ket{-\alpha})$,
$\ket{-}(\ket{\alpha}-\ket{-\alpha})$,
$\ket{+}(\ket{\alpha}+\ket{-\alpha})$, and
$\ket{+}(\ket{\alpha}-\ket{-\alpha})$, respectively.
 On the other hand, the peak at $\Delta_{\rm{q}}^{(0)}/2\pi=1.35$ MHz corresponds to a transition between the fourth excited state and the fifth excited state, which we describe by a green vertical arrow in Fig.~\ref{fig:spect1} (b),
 where the fourth (fifth) excited state is approximately given as $\ket{-}(\hat{D}_{\alpha}+\hat{D}_{-\alpha})\ket{2}$ ($\ket{-}(\hat{D}_{\alpha}-\hat{D}_{-\alpha})\ket{2}$).
 \textcolor{black}{Actually, from the numerical simulation, the population of the fourth  (fifth) excited state at $t=0$ $\mu$s 
 is $0.00590$ (0.0173) and becomes finally $0.00772$ (0.0155) at $t=3$ $\mu$s. This means that the coherent drive actually induces a transition to the fifth excited state.}
 By diagonalizing the Hamiltonian, we recognized that we have a transition from
the second excited state to the third excited state
 with an energy difference of
 $2\pi \times 0.5$ MHz.
 However, we cannot resolve this peak in the numerical simulation possibly due to the large width of the peak at $\Delta_{\rm{q}}^{(0)}$.

\begin{figure}[h]
\centering
		\includegraphics[width=9.1cm]{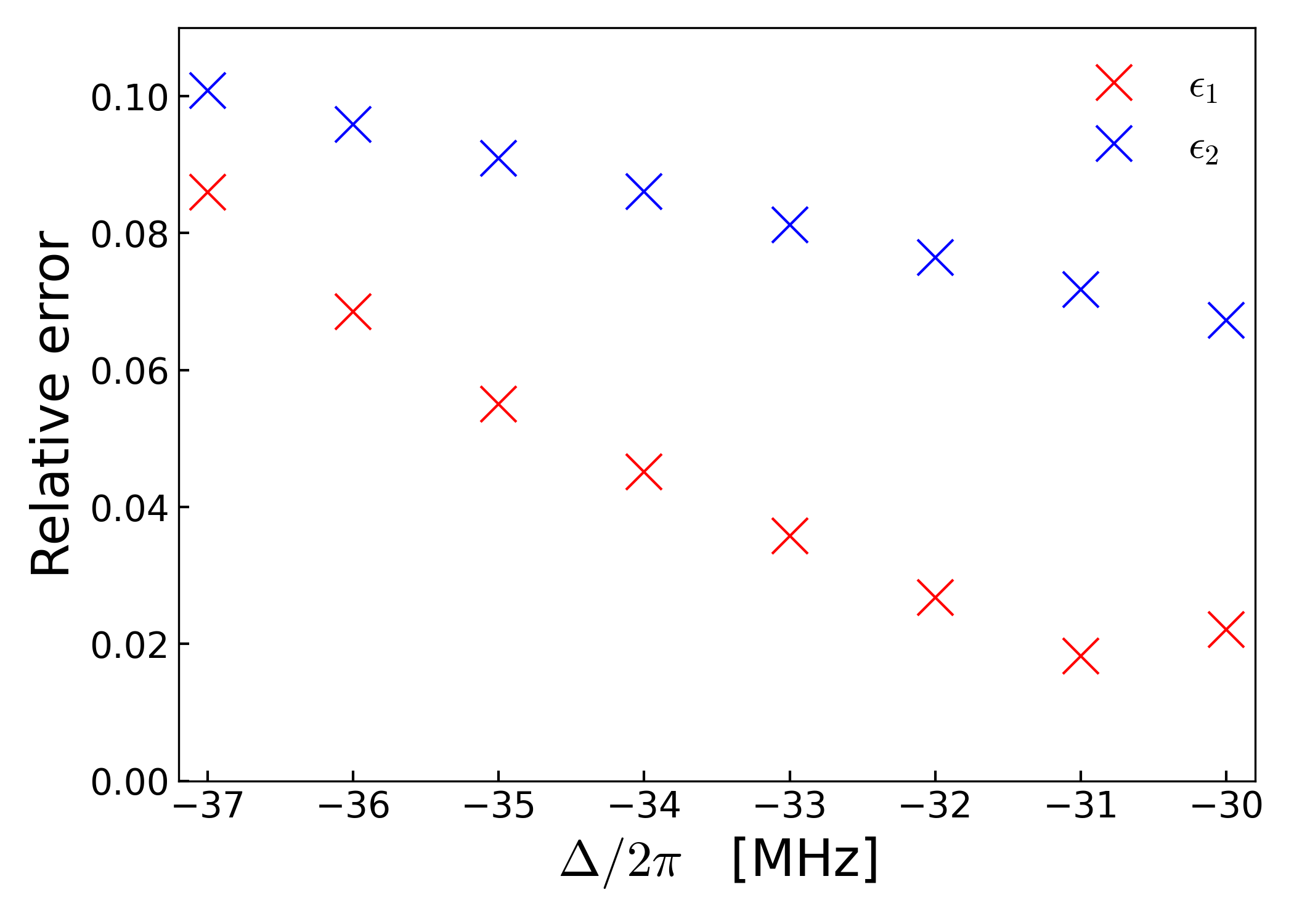} \\
\caption{
Plot of the relative error $||\alpha_{\rm{est}}|^2-|\alpha|^2|/|\alpha|^2$ 
against the detuning of the KPO
 where $|\alpha|^2$ ($|\alpha_{\rm{est}}|^2$) is the true (estimated) value of the photon number.
 We set the parameters as $\chi/2\pi=18.0$ MHz, $\beta/2\pi=42.0$ MHz, $g/2\pi=5.0$ MHz, $\lambda_{\rm{p}}/2\pi=0.5$ MHz, $\gamma_1/2\pi=\gamma_2/2\pi=0.8$ MHz, and $\omega_g=\omega_p/2$.
}
\label{fig:spec_error1}
\end{figure}

\begin{figure}[h]
\centering
		\includegraphics[width=9.1cm]{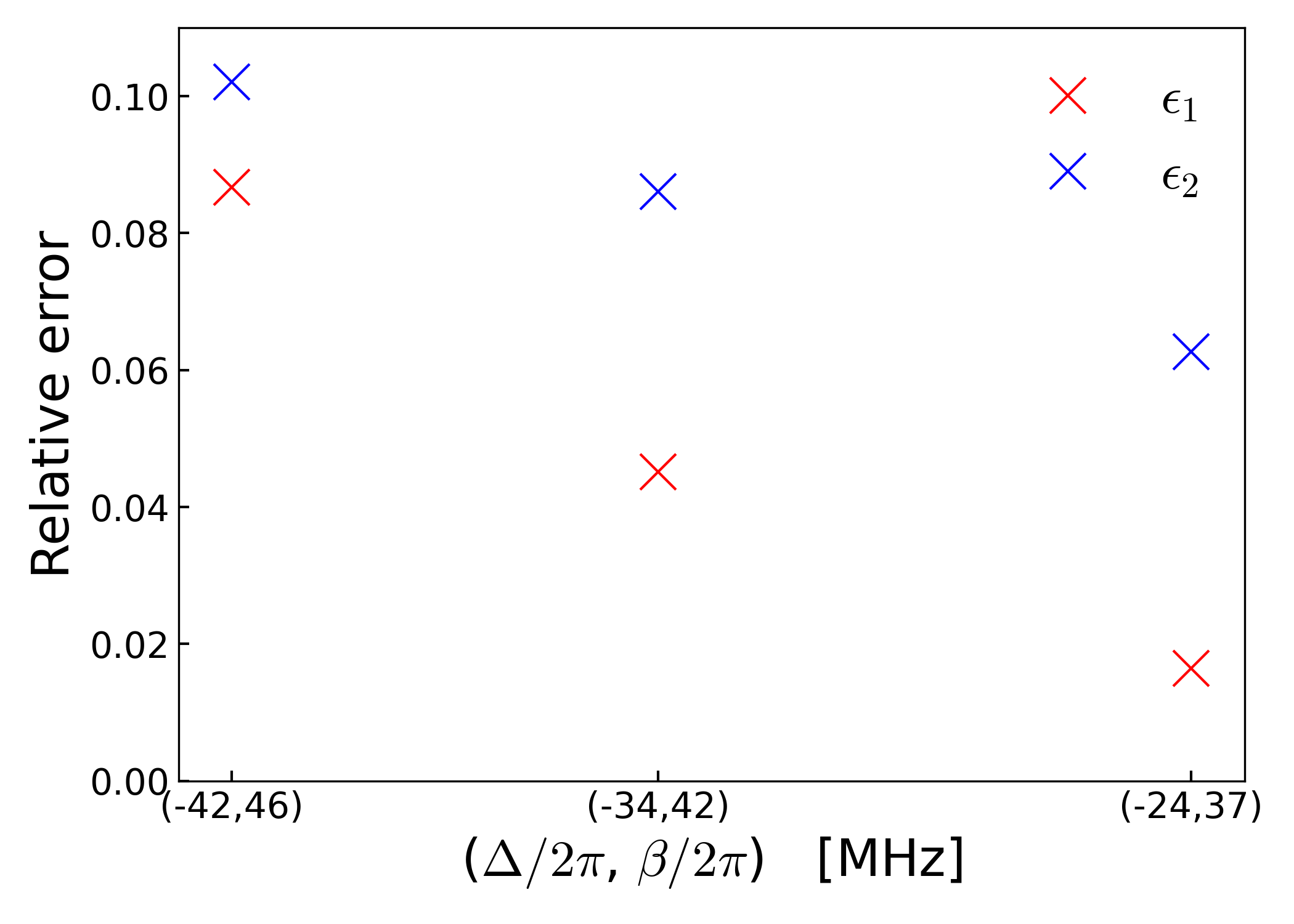} \\
\caption{
Plot of 
the relative error $||\alpha_{\rm{est}}|^2-|\alpha|^2|/|\alpha|^2$ 
against the detuning of the KPO
 where $|\alpha|^2$ ($|\alpha_{\rm{est}}|^2$) is the true (estimated) value of the photon number.
We set $\beta$ to satisfy a condition of $(2\beta + \Delta)/2\pi = 50$ MHz.
Also, we set the parameters as $\chi/2\pi=18.0$ MHz, $g/2\pi=5.0$ MHz, $\lambda_{\rm{p}}/2\pi=0.5$ MHz, $\gamma_1/2\pi=\gamma_2/2\pi=0.8$ MHz, and $\omega_g=\omega_p/2$.
}
\label{fig:spec_error2}
\end{figure}

\begin{figure}[h]
\centering
		\includegraphics[width=9.1cm]{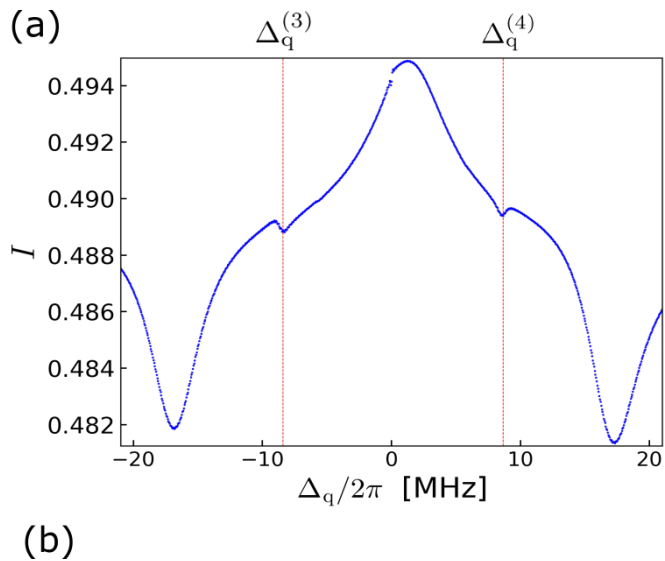} \\
\caption{
(a) The time-integrated spectra $I$ against the detuning $\Delta_{\rm{q}}/2\pi$ with the Rabi frequency $\lambda_{\rm{p}}/2\pi=2$  MHz.
We use the same parameters as those in Fig.~\ref{fig:spect1} (a) except the Rabi frequency. 
We observe not only the main two dips but also small dips
at $\Delta _{\rm{q}}^{(3)}/2\pi=-8.40$ MHz and $\Delta _{\rm{q}}^{(4)}/2\pi=-8.65$ MHz. }

\label{fig:spec2}
\end{figure}


Now, let us discuss the estimation of the number of photons.
We consider a steady state $\hrho_{\rm{ss}}$ of Eq.~\eqref{Eq.GKSL} with $\lambda_{\rm{p}}=0$ and $g=0$, and we define $\tr[\rho_{\rm{ss}}\hat{a}^\dagger\hat{a}]=|\alpha|^2$.
Let us define a
 relative error of $|\alpha|^2$ estimated by using our method as $\epsilon_1\equiv||\alpha_{\rm{est}}|^2-|\alpha|^2|/|\alpha|^2$, where $|\alpha|^2$($|\alpha_{\rm{est}}|^2$) is the actual (estimated) value of the photon number of the KPO.
Also, when we use the analytical formula, the relative error of the estimated $|\alpha|^2$ is
defined as $\epsilon_2\equiv||\alpha_{\rm{ana}}|^2-|\alpha|^2|/|\alpha|^2$.

From Fig.~\ref{fig:spect1} (a), we observe two dips at
$\Delta ^{(1)}_{\rm{q}}$ and $\Delta ^{(2)}_{\rm{q}}$ and the frequency difference is $(\Delta ^{(2)}_{\rm{q}}-\Delta ^{(1)}_{\rm{q}})/2\pi=33.90$ MHz. We can estimate the number of photons from this, as we explained before.
Since we set $g/2\pi=5$ MHz, 
we obtain an estimated value of 
$|\alpha _{\rm{est}}|^2= 2.87$,
where we solve an equation of $4g\alpha _{\rm{est}}/2\pi=(\Delta ^{(2)}_{\rm{q}}-\Delta ^{(1)}_{\rm{q}})/2\pi=33.90$ MHz.
The relative error is calculated as
$\epsilon_1=0.0280$.
On the other hand, when we use the analytical formula, we obtain $\epsilon_2=0.0672$.
This result indicates that our method provides a more accurate estimate of $\alpha$ than the conventional method. 


Also, to further quantify the performance of our method,
we calculate the relative error of our methods with other parameters, and compare the error with that of the conventional method
\footnote{
When the detuning is too large for the KPO to bifurcate, the ground state of the KPO is not the superposition of the coherent states anymore. Thus, when we plot Figs.~\ref{fig:spec_error1} and \ref{fig:spec_error2}, we choose a range of detuning for the KPO to bifurcate in these numerical simulations.}.
In Fig.~\ref{fig:spec_error1}, we plot the relative error against the detuning of the KPO $\Delta$.
In Fig.~\ref{fig:spec_error2}, we plot the relative error against 
$\Delta$ by setting $\beta$ to satisfy a condition of $(2\beta + \Delta)/2\pi = 50$ MHz.
The reason why we choose this condition is that the estimated photon number $|\alpha_{\rm{ana}}|^2$ from the analytical formula is fixed in these numerical simulations.
From Figs.~\ref{fig:spec_error1} and~\ref{fig:spec_error2}, our method provides a more accurate estimate of $|\alpha|^2$ than the conventional method when there is a detuning $\Delta$. It is worth mentioning that, in the original proposal of QA with KPO \cite{goto2016bifurcation}, KPO has a finite detuning during QA. Therefore, our scheme is useful for such circumstances.

Furthermore, we investigate how a stronger Rabi frequency affects spectroscopic measurements.
We perform numerical simulations with a Rabi frequency
of
$\lambda /2\pi = 3$ MHz.
It is worth mentioning that
we observe not only the prominent two dips but also small dips
at $\Delta _{\rm{q}}^{(3)}/2\pi=-8.40$ MHz and $\Delta _{\rm{q}}^{(4)}/2\pi=8.65$ MHz, in Fig.~\ref{fig:spec2}.
We expect that these additional dips come from the violation of the rotating wave approximation, which  will be discussed in Appendix~\ref{purtabation}.


\section{Conclusion}
In conclusion, we propose an experimentally feasible method to estimate the number of photons of the KPO. We couple an ancillary qubit with the KPO, and spectroscopic measurements of the qubit let us know the number of photons of the KPO. Our results are essential to realize QA with KPOs for solving combinational optimization problems.

This work was supported by MEXT’s Leading Initiative for Excellent Young Researchers, JST PRESTO (Grant No. JPMJPR1919), Japan. This paper is partly based on the results obtained from a project, JPNP16007, commissioned by the New Energy and Industrial Technology Development Organization (NEDO), Japan.

\bibliography{matsumoto} %
\clearpage
\widetext
\begin{center}
\textbf{\large Appendix}
\end{center}
\setcounter{equation}{0}
\setcounter{figure}{0}
\setcounter{table}{0}
\setcounter{page}{1}
\makeatletter
\renewcommand{\theequation}{S\arabic{equation}}
\renewcommand{\thefigure}{S\arabic{figure}}
\renewcommand{\bibnumfmt}[1]{[S#1]}

\appendix
\section{Calculation of the transition frequencies by using
the perturbation theory}\label{purtabation}
In this appendix, to understand the violation of the rotating wave approximation in Fig.~\ref{fig:spec2}, we calculate the second-order of the
transition probability with the effective qubit Hamiltonian~Eq.\eqref{qubit_Hamiltonian}.
We consider the following Hamiltonian
\begin{equation}
    \hH=g\alpha\ \hsig_x+\lambda_{\rm{p}}\left(\hsig_+ e^{-i\Delta _{\rm{q}}t}+\hsig_- e^{i\Delta _{\rm{q}}t}\right).
\end{equation}

We can rewrite this
Hamiltonian as
\begin{align}
    \hH=g\alpha\hsig_x + \frac{\lambda_{\rm{p}}}{2} 
    \left[(\hsig_x+i\hsig_y)e^{-i\Delta _{\rm{q}} t} + (\hsig_x-i\hsig_y)e^{i\Delta _{\rm{q}} t}\right],
\end{align}
where we use $\hsig_{\pm}=(\hsig_x \pm i\hsig_y)/2$.
We change the notation from $\{\hsig_x, \hsig_y, \hsig_z\}$ to $\{\hat{Z}, \hat{Y}, \hat{X}\}$, and get
\begin{align}
    \hH=g\alpha\ \hat{Z} + \frac{\lambda_{\rm{p}}}{2}
    \left[(\hat{Z}+i\hat{Y})e^{-i\Delta _{\rm{q}} t} + (\hat{Z}-i\hat{Y})e^{i\Delta _{\rm{q}} t}\right].
\end{align}

We can rewrite the Hamiltonian as
\begin{align}
    \hH=g\alpha\ \hat{Z} + \frac{\lambda_{\rm{p}}}{2} \left(\left\{\hat{Z}+i\frac{\hsig_+^\prime-\hsig_-^\prime}{i}\right\}e^{-i\Delta _{\rm{q}} t} + \left\{\hat{Z}-i\frac{\hsig_+^\prime-\hsig_-^\prime}{i}\right\}e^{i\Delta _{\rm{q}} t}\right),
\end{align}
where we use $\hat{X}=\hsig_{+}^\prime + \hsig_{-}^\prime$, $\hat{Y}=-i(\hsig_{+}^\prime - \hsig_{-}^\prime)$.
We move to an interaction picture defined by a unitary operation of $\hU=\exp{(i\ g\alpha\ \hat{Z})}$.
The Hamiltonian in this frame is written as
\begin{align}
    \hH_{\rm{I}}(t)
    &=\frac{1}{2} \Biggl(\hat{Z}e^{-i\Delta _{\rm{q}} t}+\hsig_+^\prime e^{i(2g\alpha-\Delta _{\rm{q}}) t}-\hsig_-^\prime e^{-i(2g\alpha+\Delta _{\rm{q}}) t}\nn[5pt]
    &\hspace{80pt}+\hat{Z}e^{i\Delta _{\rm{q}} t}-\hsig_+^\prime e^{i(2g\alpha+\Delta _{\rm{q}}) t}+\hsig_-^\prime e^{i(-2g\alpha+\Delta _{\rm{q}}) t}\Biggr).
\end{align}
We solve a time-dependent Schrodinger equation in the interaction picture as
\begin{equation}
    i\frac{d}{dt}\ket{\psi(t)}=\lambda_{\rm{p}}\hH_{\rm{I}}(t)\ket{\psi(t)}.
\end{equation}
{By performing a perturbative expansion up to the second order, we obtain 
\begin{align}
    \ket{\psi(t)}\simeq\ket{\psi(0)} 
    + \lambda_{\rm{p}}D_1\ket{\psi(0)} 
    +\lambda_{\rm{p}}^2 D_2\ket{\psi(0)}
\end{align}
\begin{align}
    D_1 & \equiv -i\int_{0}^{t} \hH_{\rm{I}}(t_1)\\[10pt]
    D_2 &\equiv (-i)^2\int_{0}^{t}\int_{0}^{t_1}dt_2\hH_{\rm{I}}(t_1)\hH_{\rm{I}}(t_2)
\end{align}
We calculate a transition probability from the initial state $\ket{\psi(0)}=\ket{\psi_0}$ to the final state $\ket{\psi_f}$ as follows.
\begin{align}
    |\braket{\psi_{f}|\psi(t)}|^2\simeq|\braket{\psi_f|\psi_0} + C^{(1)}(t) 
    +C^{(2)}(t)|^2,
\end{align}
\begin{align}
    C^{(1)}(t) &= \lambda_{\rm{p}}\braket{\psi_f|D_1|\psi_0}\\[10pt]
    C^{(2)}(t) &= \lambda_{\rm{p}}^2\braket{\psi_f|D_2|\psi_0}
\end{align}
In the limit of large (small) $\lambda_{\rm{p}}$ ($t$) by fixing a value of $\lambda_{\rm{p}}t$, 
 we can calculate the first (second) order transition probability $|C^{(1)}|^2$ ($|C^{(2)}|^2$), and obtain as the following
\begin{align}
    |\braket{\psi_{f}|\psi(t)}|^2
    &\simeq |\braket{\psi_{f}|\psi_0}|^2+A_1\ \delta({\Delta _{\rm{q}}})
    +A_2\ \delta{(2g\alpha+\Delta _{\rm{q}})}
    +A_3\ \delta{(2g\alpha-\Delta _{\rm{q}})}\\[10pt]
    &
    +A_4\ \delta{(g\alpha+\Delta _{\rm{q}})}
    +A_5\ \delta{(g\alpha-\Delta _{\rm{q}})},
\end{align}
where $A_i$ $(B_i)$ $(i=1,2,3,4,5)$ denotes 
coefficient determined by $\lambda_{\rm{p}}$, $g$, $\alpha$, and $\Delta_{\rm{q}}$.
This result clarifies the origin of the dips observed at 
$\Delta_{\rm{q}}^{(3)}$ and $\Delta_{\rm{q}}^{(4)}$ in Figs.~\ref{fig:spec2}.
}

\end{document}